\documentclass[pra,twocolumn,longbibliography]{revtex4}
\usepackage[utf8]{inputenc}
\usepackage{bm}
\usepackage{mathtools}
\usepackage{bbold}
\usepackage{amsmath}
\usepackage{amssymb}
\usepackage{amsfonts}
\usepackage[colorlinks, linkcolor=black, anchorcolor=blue,citecolor=blue,urlcolor=blue]{hyperref}
\usepackage[normalem]{ulem}
\usepackage{color}
\usepackage{float}
\usepackage[dvipsnames]{xcolor}
\usepackage{physics}

\begin{document}
	\author{Filippo Caleca}
	\affiliation{%
		ENS de Lyon, Universit\'e Lyon 1, CNRS, Laboratoire de Physique, F-69342 Lyon, France
	}%
	\author{Saverio Bocini}
	\affiliation{%
		ENS de Lyon, Universit\'e Lyon 1, CNRS, Laboratoire de Physique, F-69342 Lyon, France
	}%
	\author{Fabio Mezzacapo}
	\affiliation{%
		ENS de Lyon, Universit\'e Lyon 1, CNRS, Laboratoire de Physique, F-69342 Lyon, France
	}%
	\author{Tommaso Roscilde}
	\affiliation{%
		ENS de Lyon, Universit\'e Lyon 1, CNRS, Laboratoire de Physique, F-69342 Lyon, France
	}%
	
	\title{Giant number-parity effect and scalable spin squeezing in Luttinger liquids}
	
	\begin{abstract}
		Finite-size quantum spin systems can be magnetized by the application of a symmetry-breaking field, but in general their symmetry is expected to be restored once the field is turned off adiabatically. Recently (F. Caleca et al., arXiv:2412.15493) we have shown that systems of half-integer spins with an odd number of sites and a parity-preserving Hamiltonian can retain a finite magnetization, hence exhibiting spontaneous symmetry breaking (SSB) at finite size. Here we generalize this phenomenon to spin chains whose low-energy physics (in zero field) realizes a Luttinger-liquid phase. We observe that odd-sized chains can exhibit a phenomenon of finite-size quasi-SSB, in which a net sub-extensive magnetization, $M \sim N^{1-1/(4K)}$ is retained, where $N$ is the number of sites and $K$ the Luttinger exponent. Interestingly, the states prepared by turning off the symmetry-breaking field quasi-adiabatically display scalable spin squeezing -- namely stronger the bigger the system -- regardless of the parity of $N$. The scaling of the squeezing parameter is dictated again by the Luttinger exponent, $\xi_R^2 \sim N^{-1+1/(2K)}$. This result shows that scalable quantum correlations with metrological significance, associated typically with high-dimensional systems, can be found as well in gapless one-dimensional ones; and they are a direct consequence of the critical nature of Luttinger liquids. 	
	\end{abstract}
	
	\maketitle

	\section{Introduction} 
	
	One-dimensional quantum systems are intrinsically different from their higher-dimensional counterparts \cite{giamarchi}. In particular, for quantum spin chains with continuous symmetry and short-ranged interactions the paradigm of long-range magnetic order does not apply. 
	Instead the low-energy physics is effectively captured by a universal field theory, the Tomonaga-Luttinger-liquid (TLL) theory \cite{giamarchi}, which predicts that quantum fluctuations are sufficiently strong so as to prevent the spontaneous breaking of a continuous symmetry in the thermodynamic limit. Accordingly, spin-spin correlations decay algebraically with the distance even in the ground state, marking the appearance of quasi-long-range order (QLRO).
	
	The investigation of one-dimensional quantum systems has received a very important stimulus from quantum simulators \cite{Georgescuetal2014}. In condensed matter, spin chains are always embedded in a three-dimensional material, namely they are coupled together by residual couplings in a three-dimensional network. On the other hand, in quantum-simulation platforms such as superconducting circuits \cite{Morvanetal2022},  trapped ions \cite{Monroeetal2021}, Rydberg atoms \cite{Emperauger_2025}, or neutral atoms in optical lattices \cite{Weietal2022}, artificial spins can be arranged in perfectly one-dimensional geometries, with open or even periodic boundary conditions. Even in two- or three-dimensional geometries the spin chains can be systematically decoupled, e.g., by using deep optical lattices \cite{Guoetal2024}.  Moreover state-of-the-art platforms allow for the individual addressing of the spins, and the direct reconstruction of their correlations at equilibrium and away from it \cite{Monroeetal2021,Weietal2022,Emperauger_2025}. Hence fundamental phenomena of one-dimensional quantum physics can now be tested in a highly controlled manner. This high level of control comes of course at the cost of a drastic reduction of the system size, as compared to condensed-matter systems with Avogadro numbers of particles.  Yet, exquisite control on the number of spins can be achieved, allowing for a systematic study of finite-size scaling properties.  
	
	Inspired by the potential offered by current quantum simulation platforms, in this work we explore the fundamental effect of \emph{number parity} on the physics of quantum spin chains. Focusing on spin-$1/2$ chains, we show how, in finite systems with an odd number $N$ of particles, exact spectral degeneracy can lead to ground states exhibiting a \emph{finite} magnetization. In dimensions $d>1$, or in the presence of interactions with a sufficiently long range, this observation corresponds to spontaneous symmetry breaking (SSB) at finite size, as we recently discussed in a previous work \cite{caleca_2024}. Here, focusing on spin chains with short-ranged interactions, we show that the residual magnetization of odd-sized chains grows only sub-extensively with $N$, so that the magnetization per spin disappears in the thermodynamic limit, where SSB is not allowed in 1d. In turn, a form of \emph{quasi-}SSB is realized. The (quasi-)adiabatic preparation of states with quasi-SSB via magnetic-field ramps is analyzed, as well as their entanglement content. In particular, we numerically demonstrate that these states exhibit scalable multi-partite entanglement in the form of spin squeezing. The latter property is also shared by even-sized chains for a finite applied field, i.e., it is a robust feature of Luttinger liquids exposed to weak symmetry-breaking fields. 
	
	The paper is organized as follows: in Sec.~\ref{sec:models} we introduce the models of interest for this work and recall fundamental elements of TLL physics. In Sec.~\ref{sec:SSB},  we show how Kramers' degeneracy can lead to ground states with subextensive magnetization at finite size. In Sec.~\ref{sec:dynamic_preparation}  we study the quasi-adiabatic preparation of these states, obtained by slowly switching off a field coupled to the order parameter.  In Sec.~\ref{sec:squeezing} we move to the entanglement properties of these states, showing that, for both odd and even sizes, arbitrarily large multipartite entanglement can appear at low fields, in the form of scalable spin squeezing. In particular, we show that the scaling exponent of the squeezing parameter depends only on the Luttinger parameter $K$, thus providing a simple condition to obtain scalable squeezing in Luttinger liquids. Finally, Sec.~\ref{sec:conclusions} offers conclusions and an outlook of our work.
	
	\section{Models} \label{sec:models}
	In the following, we will consider the $U(1)$-symmetric XXZ model with interactions decaying as a power-law of the distance, whose Hamiltonian reads
	\begin{equation}
		H = - \mathcal{J} \sum_{i<j} \frac{1}{r^{\alpha}_{ij}} \bigg( S^x_i S^x_j + 
		S^y_i S^y_j + \Delta S^z_i S^z_j \bigg)~.
		\label{eq:H}
	\end{equation}
	Here $S^\mu_i$ ($\mu=x,y,z$) are spin-$1/2$ operators attached to the sites $i$ ($i=1,\dots,N$) of a one-dimensional lattice with periodic boundary conditions.	
	In the following we shall focus on the case of ferromagnetic XY interactions, $\mathcal{J}>0$. 
	The ground-state physics of this model was studied in several recent works \cite{PhysRevLett.119.023001,Schneideretal2022,Guptaetal2025}. For a sufficiently small $\alpha$, dependent on the anisotropy $\Delta$, it can exhibit long-range XY ferromagnetism; while a large $|\Delta|$ can lead to Ising-like ferromagnetism/anti-ferromagnetism. The latter phenomena are clearly not specific to 1d systems. In this work, instead, we will focus exclusively on sufficiently short-ranged interactions and sufficiently small anisotropies $|\Delta|$, such that the ground state is gapless and its long-distance properties are described by TLL theory. For $\alpha = \infty$, i.e. nearest-neighbor (n.n.) interactions, this condition is met for $-1 \leq \Delta < 1$. We refer to Refs.~\cite{PhysRevLett.119.023001,Schneideretal2022} for a determination of the $\Delta$ ranges displaying TLL theory for finite $\alpha$. 
	
	Within TLL theory \cite{giamarchi}, spin operators are mapped to bosonic field operators $\phi(x)$ and $\Pi(x)$ in continuum space, satisfying canonical commutation relations $[\phi(x),\Pi(y)]=i\delta(x-y)$. The original Hamiltonian is then mapped to a quadratic field-theory Hamiltonian (up to irrelevant terms in the renormalization group sense), which reads 
	\begin{equation}
		H \approx \frac{1}{2\pi} \int dx \bigg[ \frac{u}{K}(\nabla\phi(x))^2 + uK(\pi\Pi(x))^2\bigg]~.
	\end{equation}
	Here $u$ is the sound velocity and $K$ the so-called Luttinger parameter; their values depend on the microscopic nature of spin-spin interactions. In particular, $K$ dictates the exponent for the power-law decay of correlations functions in the ground state. In fact, Luttinger liquids are known to exhibit critical correlations functions, characteristic of QLRO. For large distances $r_{ij}$ one has \cite{giamarchi}
	\begin{equation}
		\langle S^x_i S^x_j \rangle \approx
		A \ \bigg(\frac{1}{r_{ij}}\bigg)^{1/2K} + B \ (-1)^{r_{ij}} \bigg(\frac{1}{r_{ij}}\bigg)^{2K+1/2K} + ... \label{eq:Cxx}
	\end{equation}
	where $A, B$ are non-universal amplitudes, and the omitted terms are subdominant corrections decaying faster with the distance. For the case of n.n. interactions ($\alpha=\infty$) an exact expression for $K$ is available through Bethe-ansatz \cite{giamarchi}
	\begin{equation}
		K = \frac{\pi}{2\arccos(\Delta)} ~.
		\label{eq:Bethe_K}
	\end{equation}
	For generic, non-integrable models $K$ is computed numerically by analyzing the long-range behavior of correlations functions. In the following we will mainly focus on three different models: (i) the n.n. interacting (i.e. $\alpha=\infty$) chain with antiferromagnetic interactions for the $S^z$ spin components, $\Delta=-1/2$, (ii) the XX chain with n.n. interactions ($\Delta=0$); and (iii) the dipolar XX chain ($\alpha=3$, $\Delta=0$). From the Bethe-ansatz expression of Eq.~\eqref{eq:Bethe_K} one finds that the models (i) and (ii) have Luttinger parameters taking values $K=0.75$ and $K=1$ respectively. Numerical and perturbation-theory techniques give a value $K\simeq1.7$ for the dipolar XX chain, model (iii) \cite{Guptaetal2025,Emperauger_2025}.
	
	\section{Symmetry-broken ground states}
	\label{sec:SSB}
	
	\subsection{Double degeneracy for odd $N$}
	Hamiltonians $H$ with interactions involving an even number of spins, such as Eq.~\eqref{eq:H} containing only pairwise interactions, are symmetric under time reversal $\Theta$. Spin operators are angular momenta, and they are odd under time reversal, $\Theta S^\mu_i \Theta^\dagger = -S^\mu_i$ $(\mu = x,y,z)$; but products of an even number of spin operators are instead  time-reversal invariant. This symmetry of $H$ has important consequences on its eigenstates, and in particular on its ground state, when the number $N$ of spins is odd. An odd number of $S=1/2$ spins has a total spin ${\bm J} = \sum_i {\bm S} _i$ which only admits half-integer values, i.e. ${\bm J}^2$ has eigenvalues $J(J+1)$ with $J = (2n+1)/2$ and $n \in \mathbb{N}$. 
	This means that, for $N$ odd, any Hamiltonian eigenstate $|\psi_k\rangle$ with energy $E_k$ overlaps only with  eigenvectors of ${\bm J}^2$ with half-integer total spin, $|\psi_k \rangle = \sum_{J=1/2}^{N/2} \sum_M \sum_{\lambda} \psi_{J M \lambda}^{(k)} |J, M, \lambda\rangle$, where $M = -J,.., J$ is the eigenvalue of the $J^z$ operator, and  $\lambda$ represents a set of further quantum numbers that identify uniquely the states in the decomposition. A fundamental property of the states of half-integer spins is that they cannot be invariant under time reversal, because they necessarily exhibit a net magnetization along one direction of space. If a state $|M_\mu \rangle$ has $S^\mu |M_\mu \rangle = M |M_\mu \rangle$ for some direction $\mu$, then $ \Theta S^\mu \Theta^\dagger \Theta |M_\mu \rangle = - S^\mu \Theta |M_\mu\rangle = M \Theta |M_\mu\rangle$, namely $\Theta |M_\mu\rangle$ has opposite magnetization to $|M_\mu\rangle$, and it is orthogonal to it, $\Theta |M_\mu \rangle = (-1)^{J+M} |-M_\mu \rangle$  \cite{tasaki_book}.  The state $\Theta |\psi_k \rangle =  \sum_{J, M, \lambda}  (-1)^{J+M}  \left ( \psi_{J M \lambda}^{(k)} \right )^* |J, -M, \lambda\rangle$ differs therefore fundamentally from the state $|\psi_k \rangle$, namely  $\Theta |\psi_k \rangle \neq e^{i\phi} |\psi_k \rangle$; and yet both states are eigenstates at the same energy $E_k$, since $\Theta  H \Theta^\dagger = H$. This implies that $H$ admits pairs of degenerate eigenstates \cite{caleca_2024}. This is just a restatement of Kramers' theorem \cite{tasaki_book} for spin systems whose collective spin length is not well defined, but it can only be half-integer. 
	
	In particular, in the case of the U(1) symmetric Hamiltonian in  Eq.~\eqref{eq:H} one has that $J^z$ is a good quantum number, so that Kramers pairs are simply pairs of states with opposite $J^z$. As already stated above, our focus is on ferromagnetic XY interactions leading to a gapless spectrum above a ground state with dominant ferromagnetic XY correlations. Such a ground state minimizes the quantum number $J^z$, so as to have the largest projection of the spins onto the $xy$ plane. This implies that, for odd $N$, the ground state manifold is spanned by a degenerate doublet with $J^z = \pm 1/2$, which we shall indicate as $\ket{\Psi_{\pm1/2}}$ in the following. 
	The degenerate ground states represent the bottom of the so-called Anderson tower of states \cite{PhysRev.86.694,Beekman_2019,Tasaki2019}, corresponding to the ground states in each $J^z$ sector of the spectrum.

	\subsection{Parity eigenstates with broken symmetry}
	
	Defining the spin-parity operator $P^x=\prod_{i} (2 S^x_i)$, one can notice that this commutes with the Hamiltonian in Eq.~\eqref{eq:H} but not with the $J^z$ operator. For this reason, $\ket{\Psi_{\pm1/2}}$ are not eigenstates of $P^x$, but they can nonetheless be superposed to give parity eigenvectors as $\ket{\pm}=(\ket{\Psi_{+1/2}}\pm\ket{\Psi_{-1/2}})/\sqrt{2}$, where $P^x\ket{\pm}=\pm\ket{\pm}$. 
	
	
	Here we shall prove that, in spin chains realizing a Luttinger liquid phase in their ground state, the states with well-defined parity exhibit a finite, yet subextensive magnetization, $|\langle J^x \rangle| \sim O(N^\beta)$ with $0 < \beta<1$. 
	
	The above result can be proven by establishing bounds on the modulus of the average magnetization $|\langle J^x \rangle|$. In the following we shall focus on the $|+\rangle$ state, the magnetization of the $|- \rangle$ state being simply opposite in sign. 
	
	To establish the bounds, we decompose the state on the joint eigenbasis of the commuting operators ${\bm J}^2$, $J^z$ operators, $\ket{J,M,\lambda}$. The decomposition reads
	\begin{equation}
		\ket{+} = \sum_{J=1/2}^{N/2}\sum_\lambda c_{J,\lambda} \big(\ket{J,+1/2,\lambda} + \ket{J,-1/2,\lambda} \big)
	\end{equation}
	where we made explicit the symmetry of the state under inversion of the sign of $J^z$ (which is operated by the parity operator $P^x$). 
	Using this decomposition, the average values of $J^x$ and $(J^x)^2$ read
	\begin{equation}
		\begin{split}
			\langle J^x\rangle_+ = \frac{1}{2} &\sum_J p_J \sqrt{J(J+1)-1/4} > \frac{1}{2} \sum_J p_J J\\
			\langle (J^x)^2 \rangle_+ = &\sum_J p_J [J(J+1)-1/4] \leq 2\sum_J p_J J^2
		\end{split}
	\end{equation}
	where $\langle\dots\rangle_+ = \bra{+}\dots\ket{+}$, $p_J=\sum_\lambda|c_{J,\lambda}|^2$, and we used the fact that $J\geq1/2$. 
	Since $J\leq N/2$, one finds
	\begin{equation}
		\langle J^x \rangle_+ > \frac{ \langle (J^x)^2\rangle _+}{2N} \label{eq:Jx_g_Jx2}
	\end{equation}

	\noindent Now, assuming QLRO as in TLL theory, the integration of  the correlation function in Eq.~\eqref{eq:Cxx} gives
	$$\langle (J^x)^2 \rangle = \sum_{ij} \langle S_i^x S_j^x \rangle \approx a N^{2-1/2K},$$
	where $a$ is a positive coefficient.
	Combining this with Eq.~\eqref{eq:Jx_g_Jx2} leads to
	\begin{equation}
		\langle J^x \rangle_+ > \frac{a}{2} N^{1-1/2K},
	\end{equation}
	which establishes a lower bound for the scaling of the residual magnetization.
	
	\noindent At the same time, the positivity of the variance of $J^x$ implies 
	\begin{equation}
		\langle J^x \rangle_+ \leq \sqrt{ \langle (J^x)^2 \rangle } \approx b N^{1-1/4K},
	\end{equation}
	with $b$ a second positive coefficient. This finally gives
	\begin{equation}
		\frac{a}{2}N^{1-1/2K} \leq \langle J^x \rangle_+ \leq b N^{1-1/4K}.
		\label{eq:Jx_bound}
	\end{equation}
	Hence, we see that, in quantum spin chains possessing a Luttinger-liquid ground state, for odd $N$ one can realize a ground state with a finite order parameter for XY ferromagnetism. Yet the order parameter is \emph{not} a macroscopic one, unlike what can be observed in the case of true long-range order, corresponding to the limit $K \to \infty$ \cite{caleca_2024}. 
	
	It is worthwhile discussing this result in the context of the phenomenon of spontaneous symmetry breaking (SSB). The latter is rigorously defined \cite{Palmer1982} as the persistence of a finite magnetization (per spin) after an applied field $\Omega$ which induced the magnetization is switched off, namely when $\lim_{\Omega\to 0} \langle J^x \rangle_\Omega/N = m^x \neq 0$, where $\langle ... \rangle_{\Omega}$ is the average over a state obtained dynamically or adiabatically in a finite field. In its standard definition, this phenomenon requires taking as well the limit $N\to\infty$ \emph{before} the limit $\Omega \to 0$. In Ref.~\cite{caleca_2024} we showed that systems exhibiting long-range order in the ground state as well as parity conservation can show SSB for a finite $N$, provided that $N$ is odd. In the case of spin chains exhibiting only quasi-long-range order, the residual magnetization is only sub-extensive for $N$ odd, and we are in the presence of what can be qualified as \emph{quasi-SSB} for a finite-size system.   In the following we shall discuss how to observe quasi-SSB in a protocol of experimental relevance.

	
	\section{Quasi-adiabatic preparation} \label{sec:dynamic_preparation}
	
	A way to prepare the magnetized ground states discussed in the previous section is to apply  a field $\Omega(t)$ which couples to the $J^x$ magnetization, namely to consider  a time-dependent Hamiltonian $H'(t) = H - \Omega(t) \sum_i S_i^x$; and to ramp the field down adiabatically in time. To study this protocol, we simulate the time evolution of the coherent spin state polarized along the $x$-direction, $\ket{\Psi(t=0)}=\ket{\text{CSS}_x} = \otimes_{i=1}^N |\rightarrow_x \rangle_i$ under the time-dependent Hamiltonian $H'(t)$. 
	We study such a preparation protocol by means of matrix-product states, making use of the ITensors package \cite{itensor} for the models mentioned in Sec.~\ref{sec:models}.  
	
In the following we focus on exponential field ramps, $\Omega(t)=\Omega_0 e^{-t/\tau}$, choosing $\Omega_0=20\mathcal{J}$ and $\tau=N/10$. The rationale for scaling the ramp time $\tau$ linearly with $N$ in order to achieve adiabaticity is discussed in Appendix~\ref{app:adiabatic}. 
For a perfectly adiabatic ramp one is guaranteed to prepare the state $|+\rangle$ with this protocol: indeed, since the initial state $\ket{\text{CSS}_x}$ has a well-defined parity and $P^x$ commutes with the Hamiltonian, then parity is conserved throughout the evolution, which must end with the ground state with $P^x =1$ in zero field, namely $|+\rangle$. 

\begin{figure*}[ht]
	\includegraphics[width=0.8\textwidth]{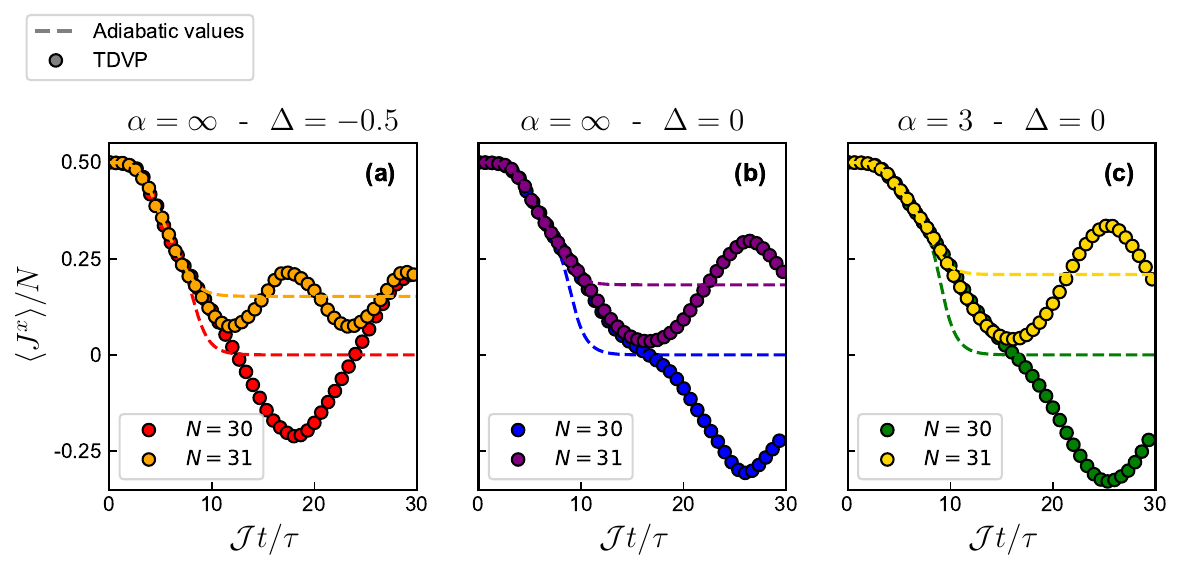}
	\caption{\emph{Giant number-parity effect in Luttinger liquids.} (a)-(c) Evolution of the magnetization along an exponential field ramp 
		$\Omega_0 e^{-t/\tau}$ for $N=30,\ 31$. Symbols correspond to the dynamical results, while the dashed line are the ground-state values for positive parity. In all panels the ramp parameters are $\tau=N/10$ and $\Omega_0=20\mathcal{J}$.}
	\label{fig:Jx_vs_t}
\end{figure*}

\subsection{Quasi-adiabatic ramps}
\label{sec:ramps}

We plot in Fig.~\ref{fig:Jx_vs_t}(a)-(c) the time-evolved magnetization as well as its adiabatic value for systems of $N=30,~31$ spins, for three different models exhibiting a Luttinger-liquid phase in their ground state. The adiabatic value corresponds to the unique ground state in the case of $N$ even; while it corresponds to the ground state  with parity $P^x = 1$ in the case of $N$ odd.  
A clear difference emerges between the even-sized and the odd-sized systems, confirming our predictions. While for even-sized systems the magnetization oscillates around zero at the end of the field ramp, for odd-sized ones it is found to oscillate around a finite value. The oscillations themselves are a consequence of the residual energy in the system above the ground states, due to the initially finite value of the field, $\Omega(0) = \Omega_0$, and to the finite duration of the ramp. 

For the slow ramps considered here the oscillations can be quantitatively understood  from the energy of the first excited state(s). For times $t\gg\tau$ the state can be approximated as overlapping uniquely with the ground state(s) and first excited states of the zero-field Hamiltonian.  In the odd-$N$ case, one has $\ket{\Psi(t)}\approx \alpha e^{-iE_{1/2}t} \ket{+} + \beta e^{-iE_{3/2}t} \ket{+_{3/2}}$, where $\ket{+_{3/2}}$ is the symmetric superposition of the lowest-excited states of the Anderson tower with $J^z = \pm 3/2$, $\ket{\Psi_{\pm3/2}}$.	
Within this assumption, one finds \cite{caleca_2024} that the oscillations should have a frequency given by $E_{3/2}-E_{1/2}$ with $E_{J^z}$ the energy of $\ket{\Psi_{\pm J^z}}$. 	
On the other hand, for an even-sized system, the state can be approximated as   $\ket{\Psi(t)}\approx \alpha e^{-iE_{0}t} \ket{\Psi_0} + \beta e^{-iE_{1}t} \ket{+_{1}}$, with $\ket{+_{1}}=(\ket{\Psi_{+1}}+\ket{\Psi_{-1}})/\sqrt{2}$. Similarly to the case of $N$ odd, one can show \cite{caleca_2024} that the oscillations possess a frequency $E_{1}-E_{0}$. 
This immediately explains the relationship between the frequencies of the oscillations for even and odd $N$. Indeed, since the energy of the states in the Anderson tower can be approximated as $E_{J^z}\simeq (J^z)^2/2I$ \cite{PhysRev.86.694,Beekman_2019,Tasaki2019}, where $I$ is a moment of inertia to be specified below, one has $E_{1}-E_{0} \simeq 1/2I$ while $E_{3/2}-E_{1/2}=1/I$, namely a $1:2$ ratio of the frequencies between even and odd $N$. 	 

Moreover, Fig.~\ref{fig:Jx_vs_N}  also shows that  the frequency of the oscillations depends strongly on the mode of interest. This can be easily understood by looking at the estimate for the moment of inertia $I$ from rotor+spin-wave (RSW) theory \cite{Roscilde_2023,PhysRevLett.131.160403}
\begin{equation}
	\frac{1}{2I} \approx \frac{\mathcal{J} \gamma_0(1-\Delta)}{2N}
	\label{eq:I}
\end{equation}
with $\gamma_0=(1/N)\sum_{i\neq j}\mathcal{J}/r^\alpha_{ij}$.
The longer range of interactions in the dipolar case causes a larger value of $\gamma_0$, which is reflected in a higher frequency of the oscillations w.r.t. the nearest-neighbor interacting XX chain. At the same time, the latter and the XXZ chain with $\Delta=-0.5$ exhibit oscillations frequencies differing by a factor $\approx 3/2$, a consequence of the $(1-\Delta)$ factor in Eq.~\eqref{eq:I}.

\subsection{Scaling of the residual magnetization} 

\begin{figure*}
	\includegraphics[width=0.8\textwidth]{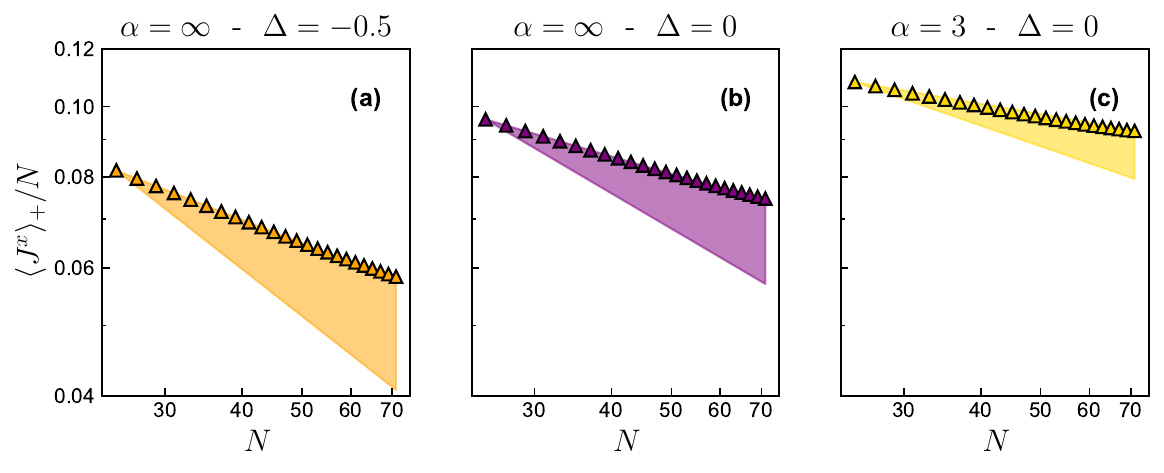}
	\caption{\emph{Giant number-parity effect in Luttinger liquids.} (a)-(c)
		Residual magnetization in odd-sized lattices as a function of the system size; shaded regions correspond to the upper and lower bounds defined by Eq.~\eqref{eq:Jx_bound}.}
	\label{fig:Jx_vs_N}
\end{figure*}

We now consider the study of the scaling of the residual magnetization as a function of the system size $N$ for odd $N$. We consider system sizes ranging from $N=25$ to $N=71$, and we  compute the positive-parity ground state by means of the Density Matrix Renormalization Group (DMRG). Results are plotted in Fig.~\ref{fig:Jx_vs_N}(a)-(c). There we find that the \emph{upper} bound given by Eq.~\eqref{eq:Jx_bound} is always almost saturated, namely the residual magnetization is found to scale as $N^{1-1/4K}$.

This result is in full agreement with what can be calculated exactly in the case of the XX chain ($\alpha = \infty, \Delta = 0$). That model can be mapped onto free fermions, and this mapping allows in turn for an exact calculation of the residual magnetization on the symmetry-broken state, as detailed in Appendix~\ref{sec:free_fermions}. The result we obtain is
\begin{equation}
	\frac{\langle J^x \rangle}{N} = N^{-1/4}\left(
	\frac{2^{1/3}e^{1/4}}{\mathit{A}^3} + o(1)
	\right)
\end{equation}
where $A$ is the Glaisher constant $\mathit{A}=1.282...$ and $o(1)$ represents sub-leading terms.  This result is perfectly consistent with the $N^{1-1/(4K)}$ scaling of $\langle J^x \rangle_+$ when $K=1$.  
In the following discussion we shall therefore assume this scaling to be generally valid for spin chains realizing Luttinger-liquid physics. 	

\begin{figure*}
	\centering
	\includegraphics[width=.9\textwidth]{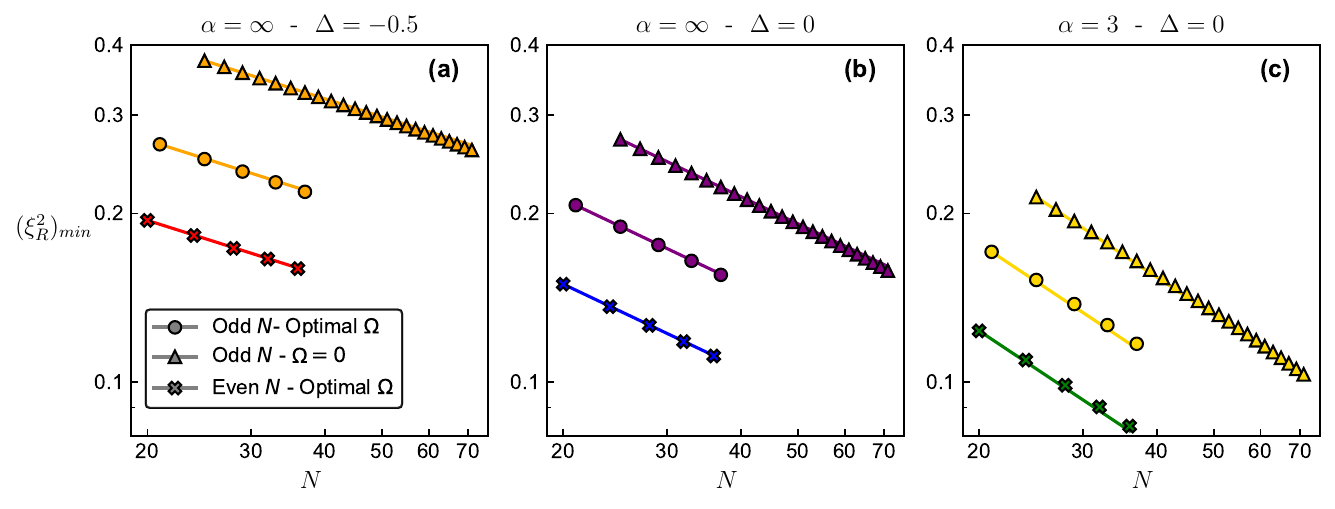}
	\caption{Scalable spin squeezing in Luttinger liquids, with different panels corresponding to the XXZ chain with $\Delta = -0.5$ (a), the XX chain (b) and the dipolar XX model (c). Triangles correspond to the spin squeezing parameter in the  $|+\rangle$ state (with odd $N$) calculated by means of DMRG after fixing the parity sector. We compare it with the minimal spin squeezing obtained in the ground state at finite $\Omega$ field, with circles (crosses) corresponding to odd (even) $N$. In all panels, solid lines correspond to the predictions of Eq.~\eqref{eq:xi_vs_N} with the $c$ prefactor adjusted so as to match the squeezing parameter for the smallest system size.} 
	\label{fig:squeezing}
\end{figure*}

\section{Scalable spin squeezing}
\label{sec:squeezing}

The states prepared via a quasi-adiabatic ramp of the $\Omega$ field possess remarkable entanglement properties for both the even-$N$ and the odd-$N$ case. 

\subsection{Odd-$N$ case}

We start the discussion with the state $|+\rangle$, which can be obtained via adiabatic ramps as discussed in the previous section. The state in question not only possesses a finite magnetization scaling as $\langle J^x \rangle_+\sim O(N^{1-1/4K})$, but it also displays minimal fluctuations of the $J^z$ collective spin component among all states with a half-integer collective spin, namely $\text{Var}(J^z)=1/4$.
The coexistence of these two aspects implies that the state exhibits \emph{spin squeezing}, as witnessed by the squeezing parameter \cite{Winelandetal1994}:
\begin{equation}
	\xi_R^2 = \frac{N\min_\theta\text{Var}(J^\theta)}{\langle J^x \rangle^2} 
\end{equation}
where $J^\theta= J^z \cos\theta + J^y \sin\theta $ is a generic collective-spin component transverse to the $x$ axis. A squeezing parameter $\xi_R^2<1$ signals that the state is entangled \cite{Sorensenetal2001} and that its entanglement makes it more sensitive to rotations (around the axis perpendicular to $J^x$ and to $J^\theta$ minimizing  $\text{Var}(J^\theta)$) than any factorized state \cite{Winelandetal1994, Pezz__2018}. More quantitatively, the condition $\xi_R^2<1/k$ (with $k$ integer) certifies an entanglement depth of at least $k+1$ spins, namely (at least) $(k+1)$-partite entanglement \cite{Pezz__2018}. 
Remarkably, the state $\ket{+}$ -- for which $\min_\theta\text{Var}(J^\theta)=\text{Var}(J^z)$ -- displays \emph{scalable squeezing}, namely a squeezing parameter decreasing with system size as
\begin{equation}
	\xi_R^2 \approx c N^{-1+1/2K} \label{eq:xi_vs_N}
\end{equation}
(with $c$ a positive prefactor) provided that $K > 1/2$ . Hence the entanglement depth in the state increases with increasing system size, albeit non-extensively. Moreover our result also defines the condition on Luttinger liquids realized by quantum spin chains to exhibit scalable spin squeezing. {Interestingly, in Ref.~\cite{Comparinetal2022} some of us observed the absence of scalability in the case $K=1/2$ (corresponding to Heisenberg or XXX chains with nearest-neighbor interactions). This observation is fully coherent with the above result. }

It is important to notice that,  approaching the long-range ordered regime ($K\rightarrow\infty$), one recovers the Heisenberg scaling $\xi_R^2 \sim N^{-1}$, i.e. the fastest possible scaling for squeezing and the related entanglement depth. This is fully consistent with what was found in our recent work on long-range ordered systems  \cite{caleca_2024}. 

We numerically confirm the prediction of Eq.~\eqref{eq:xi_vs_N} for odd-sized systems ranging from $N=25$ to $N=71$, by means of DMRG targeting the ground state in the positive parity sector.
Results are shown in Fig.~\ref{fig:squeezing}(a) for the three models already considered in Fig.~\ref{fig:Jx_vs_N}. 
We observe that the prediction of Eq.~\eqref{eq:xi_vs_N} is well verified. We stress that in Fig.~\ref{fig:squeezing}(a) the lines do not correspond to fits, but to the prediction of Eq.~\eqref{eq:xi_vs_N}, with $c$ fixed so as the reproduce the numerical value for the smallest system size.

\begin{figure*}
	\centering
	\includegraphics[width=0.8\textwidth]{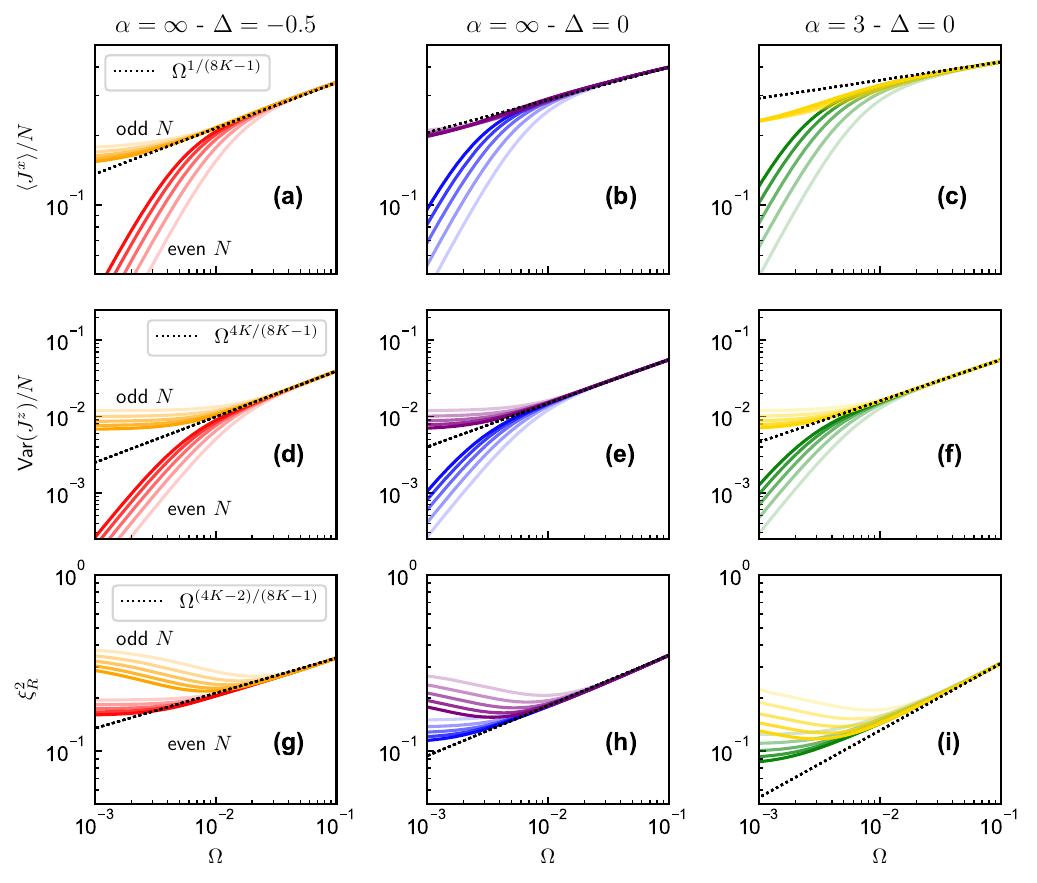}
	\caption{Scalings of (a)-(c) magnetization $\langle J^x\rangle/N$, (d)-(f) transverse variance $\text{Var}(J^z)/N$ and (g)-(i) spin squeezing parameter $\xi_R^2$ against the applied field $\Omega$. Different colors indicate different models, with the same color code used in the previous figures. Continuous lines indicate numerical results as obtained by DMRG for system sizes $N\in[20,\dots,37]$. Dotted lines represent the prediction from the theory of critical phenomena (see text).}
	\label{fig:scalings_1d}
\end{figure*}

\subsection{Even-$N$ case}

\subsubsection{Critical scaling of squeezing}
\label{sec:scaling}

In the case of odd $N$, we could predict scalable spin squeezing solely on the basis of the appearance of a residual sub-extensive magnetization in a ground state with a well-defined parity, reached after a parity-conserved field ramp. Interestingly, the same scalable squeezing can also be realized for $N$ even, provided that a \emph{finite} (yet small) field is applied to the system. 

Indeed Fig.~\ref{fig:squeezing} shows the optimal squeezing obtained in the ground state of the Hamiltonian $H'= H - \Omega J^x$ upon varying the field $\Omega$, for both $N$ odd as well as even; and the squeezing obtained for $N$ odd in the zero-field ground state $|+\rangle$ of fixed parity $P^x = 1$. There we observe that the scaling of the squeezing parameter is the same in all cases, $\xi \approx A N^{-1+1/(2K)}$ and only the overall prefactor $A$ changes. In fact the optimal squeezing for even $N$ is significantly lower than for odd $N$, something we already observed in Ref.~\cite{caleca_2024} in the case of systems with long-range ordered ground states.  


To understand the appearance of scalable spin squeezing in the even-$N$ case at finite field we need to first analyze the dependence of the two ingredients entering into the spin-squeezing parameter, $\text{Var}(J^z)$ and $J^x$, on the applied field $\Omega$. A scaling behavior of $\xi_R^2$ with $\Omega$ entails scalable spin squeezing if the field $\Omega$ is scaled in turn with system size. 

As discussed in Appendix \ref{sec:lin_theory}, treating a very weak field as a perturbation gives $\text{Var}(J^z) \sim \Omega^2$ and $J^x\sim\Omega$. This implies that the spin squeezing parameter $\xi_R^2$ loses the field dependence for $\Omega \to 0$. Yet, as also discussed in App.~\ref{sec:lin_theory}, linear-response theory breaks down for a size-dependent field, namely for $\Omega \sim N^{-2+1/(4K)}$. Hence this result still leaves the door open for scalable spin squeezing at a finite field scaling with $N$. 

To predict the behavior beyond linear-response theory one can consider that Luttinger liquids for $N\to \infty$ are critical phases, so that their response to a magnetic field coupling to the order parameter is the one proper of a critical system subject to a symmetry-breaking field. The order parameter at the critical point is expected to scale with the applied field as $J^x\sim \Omega^{1/\delta}$, with $\delta$ a critical exponent which can be related to the Luttinger one using the scaling relation (for quantum critical phases \cite{Continentino-book}) $\delta = (d + z + 2 - \eta)/(d+z-2+\eta)$, where $\eta = 1/(2K)$ is the exponent dictating the power-law decay of the correlation function at the critical point as $1/r^{d+z-2 + \eta} = 1/r^{1/(2K)}$ (see Eq.~\eqref{eq:Cxx}), and we have  $d=1$ and $z=1$ for Luttinger liquids \cite{Sachdev_2011}. As a consequence one obtains $\delta=8K-1$.

The critical field dependence of $\text{Var}(J^z)$ is less straightforward to obtain. One can argue that the $\Omega$ field immediately drives the system towards a gapped phase \cite{PhysRevX.8.011032}, in which correlations acquire an exponential decay, and in particular $\langle S_i^z S^z_{i+r} \rangle \approx C(r) e^{-r/\xi}$, where $\xi$ is the correlation length and $C(r)$ corresponds to the zero-field correlations. Consequently, one has
\begin{eqnarray}
	\frac{\text{Var}(J^z)}{N} & \approx &\sum_r C(r) ~e^{-r/\xi}  =  \sum_r C(r) \left [1 - \frac{r}{\xi} + O(\xi)^2 \right ] \nonumber  \\
	& = &  - \frac{1}{\xi} \sum_{r\neq0} r C(r)  + O(\xi^{-2})
\end{eqnarray}
where we have used the fact that $\sum_r C(r) = 0$ (the zero-field variance). 
Therefore, for $\xi \to \infty$ (corresponding to the limite ($\Omega\to 0$ when $N \to \infty$) one has that  $\text{Var}(J^z)\sim \xi^{-1}$. Combining this result with the fact that $\xi\sim\Omega^{-1/y_h}$ with $y_h=2\delta/(1+\delta)$ \cite{Goldenfeldbook}, one obtains
\begin{equation}
	\text{Var}(J^z)\sim \Omega^{\frac{1}{2}+\frac{1}{2\delta}} = \Omega^{4K/(8K-1)}~.
\end{equation}

The critical field-dependence of the spin squeezing parameter of a Luttinger liquid is therefore
\begin{equation}
	\xi_R^2 \sim \Omega^{1/2-3/2\delta} = \Omega^{(4K-2)/(8K-1)}~. \label{eq:xi_vs_Om}
\end{equation}

To translate this scaling into a finite-size scaling,  it is necessary to determine the size dependence of the field at which optimal squeezing is obtained in the ground state. 
This field can be estimated as the above-cited one, $\Omega \lesssim N^{-2+1/4K}$, below which linear response theory starts to be valid, so that the squeezing parameter stops decreasing upon decreasing $\Omega$.
Combining this condition with Eq.~\eqref{eq:xi_vs_Om} gives for the optimal squeezing for even $N$ the \emph{same} scaling as for odd $N$, namely the one reported in Eq.~\eqref{eq:xi_vs_N}~. 

\subsubsection{Numerical results}

We test all the predicted field scalings by monitoring the evolution of $J^x$, $\text{Var}(J^z)$ and $\xi^2_R$ in the ground state of $H' = H - \Omega J^x$ for a finite, yet small field $\Omega$. The results, obtained again by means of the ITensor package \cite{itensor}, are plotted in Fig.~\ref{fig:scalings_1d}; the system sizes range from $N=25$ to $N=45$. 
We find a rather good agreement with the above predictions for the power-law field scalings of the quantities of interest, governed by the Luttinger exponent. Upon increasing $N$, these scalings manifest over an increasingly large field range -- both for odd as well as even $N$. 

Similar results are presented in App.~\ref{sec:xidyn} for the field scaling of the above quantities along field ramps, following the quasi-adiabatic protocol discussed in Sec.~\ref{sec:dynamic_preparation}. The ground-state predictions for the field scalings are also observed dynamically, albeit for a smaller field range. Most importantly, thanks to the choice of the size dependence of the ramp time ($\tau \sim O(N)$), the optimal squeezing is found to scale with system size in the same way as in the ground state, namely $\xi^2_R \sim N^{-\nu}$ with $\nu = 1 - 1/(2K)$. This observation implies that the ramps, while not perfectly adiabatic, are sufficiently slow to recover the scaling properties of the adiabatic (i.e., ground-state) results.

\section{Conclusions} \label{sec:conclusions}

In this work, we have shown how a phenomenon of quasi-spontaneous-symmetry-breaking (quasi-SSB) on a finite size can appear in quantum spin chains with $S=1/2$ spins, whose ground state can be described by  Tomonaga-Luttinger-liquid theory. This phenomenon relies fundamentally on Kramers degeneracy of the ground state of a quadratic spin Hamiltonian for an odd number $N$ of spins, as well as on its property of spin-parity conservation.  Finite-size quasi-SSB amounts to the preparation of ground states exhibiting a zero-field residual magnetization, which, unlike true SSB, scales only sub-extensively with the system size. In particular, we have shown that the Luttinger-liquid exponent $K$ governs the sub-extensive scaling of the residual magnetization  $\langle J^x \rangle \sim N^{-1/4K}$. This effect requires the Hamiltonian to be quadratic and hence time-reversal invariant and spin-parity conserving, a condition potentially met in a vast class of quantum simulation platforms for quantum spin chains, such as Rydberg atom arrays \cite{Emperauger_2025}, superconducting circuits \cite{Morvanetal2022} and trapped ions \cite{Monroeetal2021}. While we have only inspected $S=1/2$ models, our results  should also extend to time-reversal-invariant chains made of  half-integer spins with $S>1/2$.

Moreover, we have shown that the adiabatic preparation of ground states of spin chains with dominant ferromagnetic XY interactions, using ramps of a field applied in the $xy$ plane, can lead to states that exhibit scalable spin squeezing -- i.e. stronger the larger the system size. This effect is directly associated with finite-size quasi-SSB for odd-$N$ chains, but it is also predicted and verified for even-$N$ chains. This is a significant result, extending previous works of some us \cite{Comparinetal2022,caleca_2024}, which related the scalability of adiabatic spin squeezing to the appearance of long-range order and SSB in the ground state.  Indeed scalability of squeezing is generally observed in systems with long-range interactions or highly connected lattices \cite{Perlinetal2020, Comparin_2022, Block_2024}, and it has been conjectured to be exclusive to systems exhibiting long-range order at the energy scale proper to the dynamical preparation of the system \cite{Block_2024} --  the ground-state energy in the case of our work. Here we show that quasi-long-range order in the ground state is sufficient for scalable spin squeezing to appear; and we define the minimal condition on the power-law decay of correlations (governed by the Luttinger exponent) for scalability to be present.

\section{Acknowledgements}
We acknowledge useful discussions with P. Holdsworth. This work is supported by PEPR-q ("QubitAF" project). All numerical calculations were performed on the CBPsmn cluster at the ENS of Lyon.

\appendix

\section{Adiabaticity condition for exponential ramps}
\label{app:adiabatic}

A quantitative criterion for an adiabatic field ramp reads \cite{Albash2018RMP}
\begin{equation}
	\min_t \frac{\delta_{\Omega(t)}^2}{|\dot{\Omega}(t) ~ \bra{\psi_1(\Omega(t))} J^x \ket{\psi_0(\Omega(t))}|} \gg1 
	\label{eq:R_t}
\end{equation}
where $\ket{\psi_0(\Omega(t))}$, $\ket{\psi_1(\Omega(t))}$ are the instantaneous ground state and first excited state respectively of $H'(t) = H  - \Omega(t) J^x$,  and 
$\delta_{\Omega(t)}$ is the energy gap that separates them.  
In the case of an exponential ramp, one has $|\dot{\Omega}(t)|=\Omega(t)/\tau$, so that the above condition becomes
\begin{equation}
	\min_t \frac{\tau ~ \delta_{\Omega(t)}^2}{|\Omega(t) \bra{\psi_1(\Omega(t))} J^x \ket{\psi_0(\Omega(t))}|} \gg1.
\end{equation}
First, we need to reconstruct the field dependence of the gap $\delta_{\Omega}$.  
To do so, we can use again the fact that a Luttinger liquid is a quantum critical phase with dynamical critical exponent $z=1$. Therefore, a perturbation such as the $\Omega$ field will open a gap related directly to the induced finite correlation length (see Sec.~\ref{sec:scaling}) as 
\begin{equation}
	\delta_\Omega \sim \xi^{-z} \sim \Omega^{1/y_h}~. \label{eq:delta_vs_Omega}
\end{equation}

Now, we can assume that the minimum of the ratio in Eq.~\eqref{eq:R_t} is reached when the gap induced by the $\Omega$ field matches the gap of the Anderson tower of states, which is the residual gap in zero field -- namely when $\delta_{\Omega(t)}\sim 1/N$. Using Eq.~\eqref{eq:delta_vs_Omega}, we find that the associated field realizing this condition scales with the system size as
\begin{equation}
	\Omega_C \sim N^{-y_h} \sim N^{-(8K-1)/4K} = N^{-2+1/4K}.
\end{equation}
The adiabaticity condition consequently becomes
\begin{equation}
	\frac{\tau ~ \Omega_C^{1/(8K-1)}}{\bra{\psi_1(\Omega_C)} J^x \ket{\psi_0(\Omega_C)}} \sim \frac{\tau ~ N^{-1/4K}}{N^{1-1/4K}} =
	\frac{\tau}{N} \gg 1,
\end{equation}
justifying the choice linearly scaling the ramp time constant $\tau$ with the system size $N$, as done in Sec.~\ref{sec:dynamic_preparation}.

\section{Residual magnetization in the odd-$N$ XX chain from free fermions} \label{sec:free_fermions}

One of the models discussed in the main text is the nearest-neighbor XX chain ($\Delta = 0, \alpha = \infty$).   
This model can be mapped onto a quadratic fermionic Hamiltonian by means of the Jordan-Wigner transformation,
\begin{equation}
	2S^z_\ell = -i\, a_{2\ell-1} a_{2\ell}, 
	\qquad 
	2S^x_\ell = \left(\prod_{j=1}^{\ell-1} 2S^z_j \right) a_{2\ell-1},
\end{equation}
where the operators $a_j$ are Majorana fermions satisfying the canonical anticommutation relations 
$a_i a_j + a_j a_i = 2\delta_{i,j}$.

In this appendix, we show how to use free-fermion techniques to compute $\langle J^x\rangle_+$ in chains of odd length $N$.  
For a complete account of the free-fermion formalism employed here, we refer the reader to Ref.~\cite{bocini_thesis}.  

We start by rewriting $\langle J^x\rangle_+$ as  
\begin{multline}
	\frac{\langle J^x\rangle_+}{N} 
	= \bra{\Psi_{+1/2}} 2 S^x_N \ket{\Psi_{-1/2}} 
	\\
	= \bra{\Psi_{+1/2}} 2 S^x_N 
	\!\left(\prod_{\ell=1}^N 2S_\ell^x\right)\! 
	\ket{\Psi_{+1/2}}
	\\
	= \bra{\Psi_{+1/2}} \prod_{\ell=1}^{N-1} 2 S^x_\ell \ket{\Psi_{+1/2}}
	\\
	= \bra{\Psi_{+1/2}} 
	\prod_{\ell=1}^{\frac{N-1}{2}} (-i\, a_{4\ell-2}a_{4\ell-1}) 
	\ket{\Psi_{+1/2}},
\end{multline}
where we have used translational invariance and the relation 
$\ket{\Psi_{-1/2}} = \prod_{\ell=1}^N 2S_\ell^x \ket{\Psi_{+1/2}}$.

One can show that the state $\ket{\Psi_{+1/2}}$ is the ground state of the Ramond or Neveu–Schwarz fermionic sector \cite{bocini_thesis}, depending on whether $(L-1)/2$ is odd or even, respectively.  
In particular, $\ket{\Psi_{+1/2}}$ is a Gaussian state and, as such, is completely characterized by its two-point correlations via Wick's theorem.  
Defining
\begin{equation}
	\Gamma_{m,n} = \delta_{m,n} - \langle a_m a_n \rangle,
\end{equation}
the nonvanishing components read
\begin{equation}
	\begin{aligned}
		\Gamma_{2m+1,2n+2}
		&= -\Gamma_{2m+2,2n+1}
		= \frac{i}{L}\, \mathcal{F}(m-n),
	\end{aligned}
\end{equation}
where
\begin{equation*}
	\tiny 
	\mathcal{F}=\left\{\begin{aligned}
		\sum_{k=-\frac{-1}{4}}^{\frac{N-1}{4}}
		\cos(\frac{2\pi k m}{N})-(-1)^{m}
		\sum_{k=-\frac{N-5}{4}}^{\frac{N-1}{4}}
		\cos(\frac{\pi  m (2k-1)}{N}) \textrm{, even }\frac{N-1}{2}
		\\
		\sum_{k=-\frac{N-3}{4}}^{\frac{N-3}{4}}
		\cos(\frac{2\pi k m}{N})-(-1)^{m}
		\sum_{k=-\frac{N-3}{4}}^{\frac{N+1}{4}}
		\cos(\frac{\pi  m (2k-1)}{N}) \textrm{, odd }\frac{N-1}{2}
	\end{aligned}\right.
\end{equation*}

Since 
$\prod_{\ell=1}^{N-1} 2 S^x_\ell$ commutes with $\prod_{\ell=1}^N 2S^z_\ell$ (for odd $N$), 
the expectation value can be computed entirely within the fermionic sector to which $\ket{\Psi_{+1/2}}$ belongs.  
Using Wick’s theorem, we obtain
\begin{equation}
	\bra{\Psi_{+1/2}} \prod_{\ell=1}^{N-1} 2 S^x_\ell \ket{\Psi_{+1/2}}
	= \operatorname{Pf}(A),
\end{equation}
where $\operatorname{Pf}$ denotes the Pfaffian and 
$A$ is the antisymmetric $(N-1)\!\times\!(N-1)$ submatrix of $\Gamma$ defined as 
\begin{equation}
	A_{m,n}=i\,\Gamma_{2m-1+\operatorname{mod}_2(m),\,2n-1+\operatorname{mod}_2(n)}.
\end{equation}

Since $A_{2m+1,2n+1}=A_{2m+2,2n+2}=0$, 
the Pfaffian reduces to a determinant,
$\operatorname{Pf}(A)=\det(B)$,
where $B$ is the real $\frac{N-1}{2}\times\frac{N-1}{2}$ Toeplitz matrix defined by $B_{m,n}=A_{2m-1,2n}$.
Explicitly,
\begin{multline}
	B_{m,n}
	= \frac{1}{N}\, \mathcal{F}(2m-2n-1)
	\\
	= \frac{1}{N}\sum_{k=-\frac{N-1}{2}}^{\frac{N-1}{2}}
	\cos\!\left(\frac{\pi k (2m-2n-1)}{N}\right)
	\\
	= \frac{1}{N}\,
	\frac{\sin\!\left[\frac{\pi}{2}(2m-2n-1)\right]}
	{\sin\!\left[\frac{\pi}{2L}(2m-2n-1)\right]}.
\end{multline}
In the thermodynamic limit $N\to\infty$,
\begin{equation}
	B_{m,n}\ \longrightarrow\ 
	\operatorname{sinc}\!\left[\tfrac{\pi}{2}(2m-2n-1)\right],
\end{equation}
where $\operatorname{sinc}(x)=\sin(x)/x$.

The symbol of this Toeplitz matrix in the large-$N$ limit is
\begin{equation}
	\hat{B}(e^{ip})
	= \sum_{m\in\mathbb{Z}} B(m)e^{imp}
	= e^{i p/2}, 
	\qquad p\in(-\pi,\pi],
\end{equation}
periodically extended outside this interval.  
Note that the symbol is discontinuous at $p=(2n+1)\pi$, $n\in\mathbb{Z}$.

The asymptotic behavior of $\det(B)$ for $N\to\infty$ can then be obtained from Ref.~\cite{Krasovsky_2011}.  
In the notation of Eq.~(2.3) therein, the symbol $\hat{B}$ corresponds to the parameters
$n=(N-1)/2$, $m=1$, $V=0$, $\alpha_0=\alpha_1=\beta_0=0$, $\beta_1=1/2$, and $\theta_1=\pi$.  
Using Theorem~(2.2) of the same reference to study $N\rightarrow+\infty$, we find
\begin{equation}
	\frac{\langle J^x\rangle_+}{N} 
	= \det(B) 
	=
	N^{-1/4}
	\!\left(
	\frac{2^{1/3}e^{1/4}}{\mathit{A}^3} + o(1)
	\right),
\end{equation}
where $\mathit{A}=1.282\ldots$ is the Glaisher constant, 
and $o(1)$ denotes subleading corrections.

 \begin{figure*}[ht]
 	\centering
 	\includegraphics[width=0.8\textwidth]{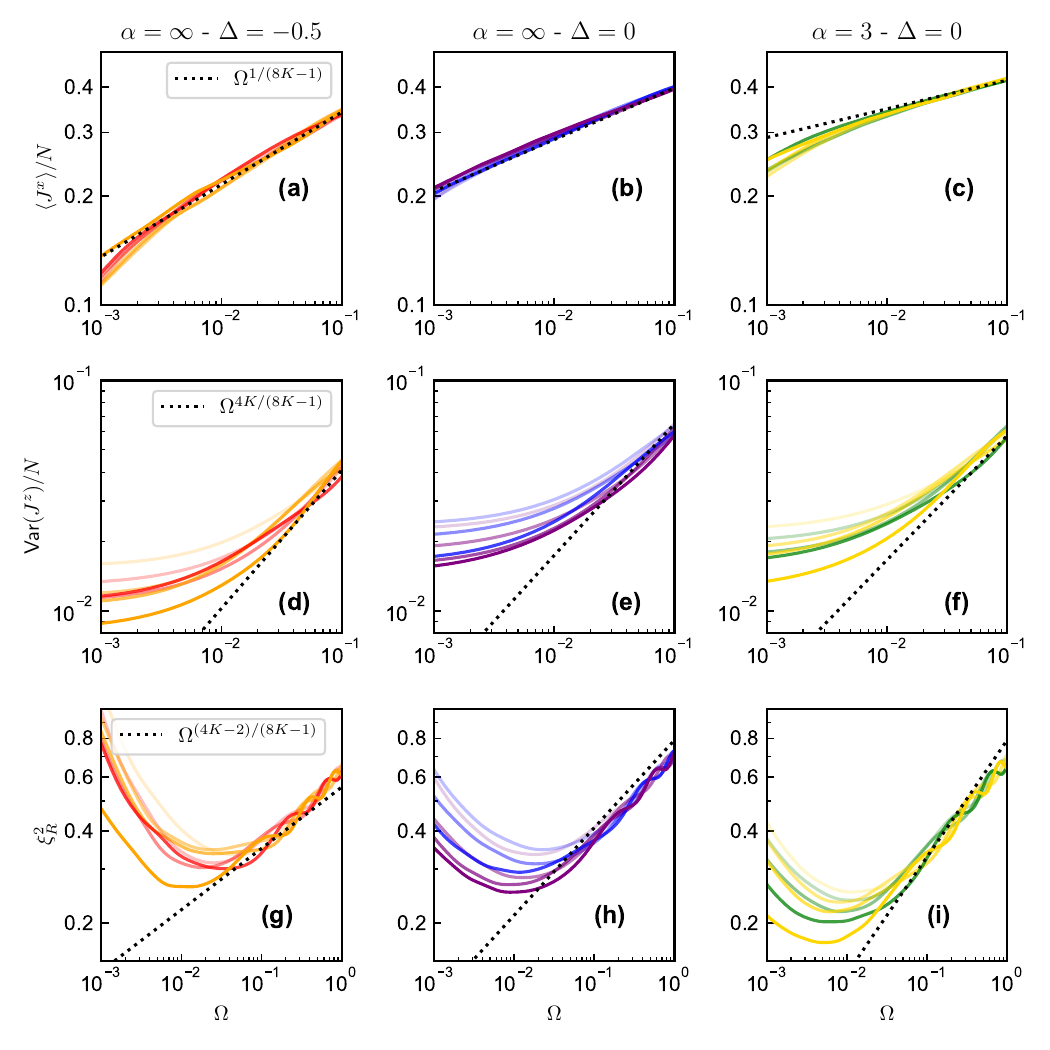}
 	\caption{Scalings of (a)-(c) magnetization $\langle J^x\rangle/N$, (d)-(f) transverse variance $\text{Var}(J^z)/N$ and (g)-(i) spin squeezing parameter $\xi_R^2$ against the $\Omega$ field as obtained along the field ramps discussed in Sec.~\ref{sec:dynamic_preparation}. Different colors indicate different models, with the same color code used in the previous figures. Continuous lines indicate numerical results for system sizes $N\in[25,\dots,41]$. Dotted lines represent the prediction from the theory of critical phenomena.}
 	\label{fig:scalings_TDVP}
 \end{figure*}

\begin{figure}
 	\centering
 	\includegraphics[width=0.45\textwidth]{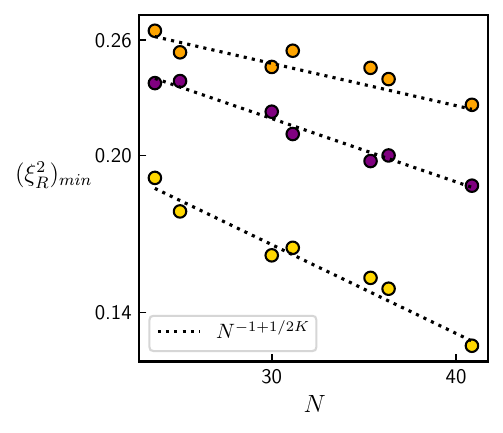}
 	\caption{Minimum of the spin squeezing parameter along the field ramps shown in Fig.~\ref{fig:scalings_TDVP}, and plotted against the system size $N$. Different colors indicate different models (same color code as in the previous figures). The dotted lines show the power-law scaling $\xi_R^2 \approx c  N^{-1+1/(2K)}$ expected for Luttinger liquids, with a prefactor  $c$ fitted to each data set.}
 	\label{fig:xi_TDVP}
\end{figure}

\section{Linear response theory} \label{sec:lin_theory}
In this section, we deal with linear response theory for the ground state of the Hamiltonian of Eq.~\eqref{eq:H} under application of a weak $\Omega$ field. 
Calling $\ket{\Psi^{(0)}_0}$ the ground state of the unperturbed model in zero field, and $E^{(0)}$ its energy, the perturbed ground state reads, at first order in perturbation theory
\begin{equation}
	\ket{\Psi^{(1)}_0} \simeq \ket{\Psi^{(0)}_0} - \Omega \sum_{k\neq0} \frac{\bra{\Psi^{(0)}_k} J^x \ket{\Psi^{(0)}_0}}{E_k^{(0)}-E_0^{(0)}}\ket{\Psi^{(0)}_k}~,
\end{equation}
where $\ket{\Psi^{(0)}_k}$, $E^{(0)}_k$ are the unperturbed $k-$th eigenstate and its corresponding energy, respectively.
Since for an even number of spins the unperturbed ground state has $J^z=0$, the sum only contains states with $J^z=\pm1$. Moreover, the sum should be dominated by the lowest-energy states with $J^z=\pm1$, i.e. the ones contained in the Anderson tower of states. In fact, the $J^x$ operator cannot create spin-waves with finite momentum on top of the Anderson tower. Calling $\ket{\Psi_{J^z=\pm1}^{(0)}}$ these two states, one finds
\begin{equation}
	\ket{\Psi^{(1)}_0} \approx  \ket{\Psi^{(0)}_0} + \Omega \sum_{m=-1,+1} 
	\frac{\bra{\Psi^{(0)}_{J^z=m}} J^x \ket{\Psi^{(0)}_0}}{1/2I_N}
	\ket{\Psi^{(0)}_{J^z=m}}.
\end{equation}
The last expression yields the following scalings
\begin{equation}
	\begin{split}
		\langle J^x \rangle & \sim \Omega \\
		\langle (J^z)^2 \rangle & \sim \Omega^2,
	\end{split}
\end{equation}
which imply $\xi_R^2\sim\Omega^0$, hence squeezing becomes independent of the applied field.
Nonetheless, for perturbation theory to be consistent, the condition
\begin{equation}
	\Omega~\frac{\bra{\Psi^{(0)}_{J^z=\pm1}} J^x \ket{\Psi^{(0)}_0}}{1/2I_N} \lesssim 1
	\label{eq:condition}
\end{equation}
must be satisfied, requiring that the correction to the unperturbed ground state is actually small. The moment of inertia $I_N$ scales linearly with $N$ while, as discussed in the main text, the matrix element 
$\bra{\Psi^{(0)}_{\pm1}} J^x \ket{\Psi^{(0)}_0}$
scales as $N^{1-1/4K}$. Consequently, perturbation theory is valid as long as Eq.~\eqref{eq:condition} is valid, which requires
\begin{equation}
	\Omega\lesssim N^{-2+1/4K}~.
\end{equation}

\section{Dynamical results for the squeezing parameter} 
\label{sec:xidyn}

To test the adiabaticity of exponential ramps with $\Omega(0)=20\mathcal{J}$ and $\tau=N/10$, and to check the experimental feasibility of our protocol, we compute the real time evolution using the time dependent variational principle (TDVP) with MPS wavefunctions. Results are plotted in Fig.~\ref{fig:scalings_TDVP}. There we observe that the power-law field scaling of the quantities of interest ($\langle J^x \rangle$, ${\rm Var}(J^z)$ and $\xi_R^2$) expected from the theory of critical phenomena is exhibited over a sizable field range, although the discrepancy with respect to the ground-state results (i.e. the perfectly adiabatic ones) of Fig.~\ref{fig:scalings_1d} is rather apparent. In particular both the odd as well as even sizes show similar behaviors. This is not surprising, since the effect of number parity on the dynamical behavior is actually observed at even smaller fields (and corresponding magnetizations $\langle J^x \rangle$) than the ones in the range of Figs.~\ref{fig:scalings_1d} and \ref{fig:scalings_TDVP}, see Fig.~\ref{fig:Jx_vs_t}. Hence one could legitimately ask what aspects of the adiabatic behavior survive in our time-dependent simulations. 

Interestingly, all squeezing curves in Fig.~\ref{fig:scalings_TDVP}(g-h-i) exhibit a clear optimum at an intermediate field; and both the optimum value as well as the optimal field appear to scale with system size. We examine the scaling of optimal squeezing in Fig.~\ref{fig:xi_TDVP} for the three models of interest, and compare it with the scaling expected for the ground-state results. 
There we observe rather convincingly that scalable spin squeezing with the same scaling exponents as in the perfectly adiabatic state is exhibited by our dynamical data. This means that the choice of ramp times $\tau = N/10$, scaling linearly with system size, is not sufficient to guarantee perfect adiabaticity (as seen both in Fig.~\ref{fig:Jx_vs_t}, as well as in the comparison of Figs.~\ref{fig:scalings_1d} and \ref{fig:scalings_TDVP}); but it is nonetheless sufficient to recover the right \emph{scaling} of optimal squeezing with system size. This means that we can observe dynamically the right scaling of entanglement properties expected in the ground state. This property can be taken as the defining feature of \emph{quasi}-adiabatic ramps. 

\bibliography{bibliography.bib}

\end{document}